Research article

# *Ab-initio* calculations of spin tunneling through an indirect barrier

Athanasios N Chantis[1], Titus Sandu*[2] and Jialei L Xu[3]

Address: [1]Theoretical Division, Los Alamos National Laboratory, Los Alamos, New Mexico 87545, USA, [2]International Centre of Biodynamics, Intrarea Portocalelor Street, Nr.1B, Postal Code 060101, District 6, Bucharest, Romania and [3]Arizona State University, Tempe, Arizona, 85284, USA

Email: Athanasios N Chantis - achantis@lanl.gov; Titus Sandu* - tsandu@biodyn.ro; Jialei L Xu - jialei.xu@asu.edu

* Corresponding author

## Abstract

We use a fully relativistic layer Green's functions approach to investigate spin-dependent tunneling through a symmetric indirect band gap barrier like GaAs/AlAs/GaAs heterostructure along [100] direction. The method is based on Linear Muffin Tin Orbitals and it is within the Density Functional Theory (DFT) in the Local Density Approximation (LDA). We find that the results of our *ab-initio* calculations are in good agreement with the predictions of our previous empirical tight binding model [Phys. Rev. **B**, 075313 (2006)]. In addition we show the $k_\parallel$-dependence of the spin polarization which we did not previously include in the model. The *ab-initio* calculations indicate a strong $k_\parallel$-dependence of the transmission and the spin polarization due to band non-parabolicity. A large window of 25–50% spin polarization was found for a barrier of 8 AlAs monolayers at $k_\parallel$ = 0.03 $2\pi/a$. Our calculations show clearly that the appearance of energy windows with significant spin polarization depends mostly on the location of transmission resonances and their corresponding zeros and not on the magnitude of the spin splitting in the barrier.

## Background

The possibility of spin-polarized transmission through a zinc-blende semiconductor symmetric barrier was investigated in a few previous articles [1-3]. In Ref. [1] only direct band gap materials were considered for the barrier. Nevertheless, many zinc-blende wide band gap semiconductors are indirect, therefore the indirect tunneling through a barrier must also be considered. This feature was addressed in Ref. [2]. The authors argued that the linear-$k$ spin-orbit splitting at $X$ point will induce larger effect than the $k^3$-term at $\Gamma$ point [2]. They found an energy window around the top of the barrier, where the effect arising from the linear-$k$ term at $X$ point is larger than the effect stemming from the $k^3$-term at $\Gamma$ point.

Mishra *et al.* made no reference to the importance of the interaction between the discrete states at the *X* valley in the barrier and the continuum [2]. In their work, the authors considered the electrons close to the top of the indirect barrier, where the indirect tunneling becomes dominant [2]. In our previous work [3] we have shown that, at these energies, the $\Gamma$-*X* mixing plays a greater role than Mishra *et al.* suggested [2]. We have also considered the Fano resonances that occur because of the interaction between the discrete states at the *X* point in the barrier and the continuum at the $\Gamma$ point in the contacts. To be more precise, in such systems an incident electron can tunnel either directly through evanescent $\Gamma$ state in the barrier or through quasi-bound *X* state in the barrier. The resonance occurs due to resonant tunneling through quasi-bound *X* state, while the anti-resonance occurs due to destructive interference between the two channels of electron transmission [4]. Below or above the resonance the two channels are out of phase. The zero in the transmission occurs whenever the amplitude magnitudes of the two channels become equal. In our previous paper [3] we have used a realistic empirical tight binding model to give a unified description of the spin dependent tunneling through an indirect symmetric barrier with the GaAs/AlAs/GaAs heterostructure as an example. We showed several interesting aspects of such process, with particular emphasis on the large energy windows of spin polarization that can be obtained when appropriate conditions are satisfied.

Although our study presented quantitative results, several simplifications were made. The tight-binding Hamiltonian assumes parabolic conduction bands and it does not distinguish between $X_1$ and $X_3$ states [3], whereas the conduction band minimum is at $\Delta$, along $\Gamma$-*X* line. Moreover, the complex structure of spin-dependent evanescent states in the band gap was completely ignored. This complicated structure of spin-dependent evanescent states might play an important role in the spin-dependent transport due to the meV scale on which the spin splittings occur [5-7]. In the present paper we extend the previous work [3] by reporting the results of the first *ab-initio* multiband study of spin-dependent tunneling through a zinc-blende barrier. The calculations are performed at the LDA level of the DFT with its well-known shortcomings regarding the band gaps and the effective masses. From the point of view of device modeling this may constitute a handicap; tools like $sp^3d^5s*$[8] full band semi-empirical tight-binding would perform a better job in reproducing band gaps, effective masses, and complex and imaginary bands. In this work we are checking if the main points made in the previous work [3] (namely, the criteria needed for obtaining large energy windows with significant spin polarization) are still valid in the context of truly multi-band calculations. The current calculations take into account the full band structure of the GaAs/AlAs/GaAs [100] heterostructure (including the spin dependency of the imaginary bands in the barrier) and, unlike our previous two-band tight-binding calculations, the *ab-initio* multiband calculations show that the nonparabolicity should be also considered in order to obtain large spin polarization.

## Results and discussion

Our approach is based on the fully-relativistic first-principles transport method. The method is based on the Green's function representation of the tight-binding linear muffin-tin orbital (TB-LMTO) basis in the atomic spheres approximation [9]. Within the relativistic formulation of the Local Spin Density Approximation (LSDA) in which only the spin component of the current density is taken into account [10], inside each atomic sphere we solve the Kohn-Sham Dirac equation [11]

$$[c\vec{a}\vec{p} + (\beta - I)mc^2 + V(r) + \mu_B B(r)\vec{n}\vec{\Sigma}]\Psi(E, \vec{r}) = E\Psi(E, \vec{r}), \tag{1}$$

where,

$$\alpha = \begin{pmatrix} 0 & \vec{\sigma} \\ \vec{\sigma} & 0 \end{pmatrix}, \beta = \begin{pmatrix} I & 0 \\ 0 & -I \end{pmatrix}, \Sigma = \begin{pmatrix} \vec{\sigma} & 0 \\ 0 & \vec{\sigma} \end{pmatrix}. \tag{2}$$

Here, $\vec{\sigma}$ is the vector of Pauli matrices, $\vec{p}$ is the momentum operator, and $\vec{n}$ is the unit vector in the direction of the effective magnetic field inside the Muffin-Tin (MT) sphere. The energy E is referenced to the total relativistic energy $W = mc^2 + E$. The effective magnetic field in equation (1) can be found as $B(r) = (V^{\uparrow}(r) + V^{\downarrow})(r)/2$ [12]. The solutions of the Kohn-Sham Dirac equation are linear combinations of bispinors:

$$\Psi_\mu(\vec{r}, E) = \sum_\kappa \Psi_{\kappa\mu}(\vec{r}, E), \tag{3}$$

$$\Psi_{\kappa\mu}(\vec{r}, E) = \begin{pmatrix} g_{\kappa\mu}(E, r)\Omega_{\kappa\mu}(r) \\ if_{\kappa\mu}(E, r)\Omega_{-\kappa\mu}(r) \end{pmatrix}. \tag{4}$$

Here, $\Omega_{\kappa\mu}(\hat{r})$ are the spin spherical harmonics, $\mu$ is the projection of the total angular momentum and $\kappa$ is the relativistic quantum number: $\kappa^2 = J(J+1) + \frac{1}{4}$.

Within the fully relativistic LMTO method the boundary conditions at the MT radius $s$ are given in the form of two matrix equations [13,14]

$$N(E) = (2l+1)\left(\frac{w}{s}\right)^{l+1} g^{-1}(E, s)(D(E, s) - Il)^{-1}, \tag{5}$$

$$P(E) = 2(2l+1)\left(\frac{w}{s}\right)^{2l+1} (D(E, s) + Il + I)(D(E, s) - Il)^{-1}, \tag{6}$$

where, $N(E)$, $P(E)$, $D(E)$ and $g(E)$ are 2 × 2 matrices for each value of $\kappa$, $\mu$ and site $R$. $N(E)$ and $P(E)$ are arbitrary matrices defined by the boundary conditions, with elements the $\kappa$, $\mu$ components of the so called potential and normalization functions defined in the scalar relativistic LMTO method [15]. $D(E, s)$ and $g(E, s)$ are matrices with elements the $\kappa$, $\mu$ components of logarithmic derivative and large component of the wave function. The primary difference of this work with previous fully relativistic formulations of LMTO method is that we use the third-order parametrization of the potential functions [16,17]: the radial amplitudes are expanded up to quadratic terms in linearization energy $\varepsilon_\nu$,

$$\Psi_{\kappa\mu}(E,\vec{r}) = \Psi^\nu_{\kappa\mu}(\vec{r}) + (E - \varepsilon_\nu)\dot{\Psi}^\nu_{\kappa\mu}(\vec{r}) + \frac{1}{2}(E - \varepsilon_\nu)^2\ddot{\Psi}^\nu_{\kappa\mu}(\vec{r}). \tag{7}$$

The Green's function of the layered system is constructed by the principal-layer technique [13]. The layers from $-\infty$ to 0 and from $N + 1$ to $\infty$ are the contacts, while the layers from 1 to $N$ are the active layers. Then, the transmission coefficient within the Landauer-Büttiker approach [18] can be calculated as

$$T(k_{||}, E) = \lim_{\varepsilon \to 0^+} \frac{1}{2} Tr[B_1(k_{||}, E)g_{1,N}(k_{||}, z_+)B_N(k_{||}, E)g_{N,1}(k_{||}, z_-) + B_1(k_{||}, E)g_{1,N}(k_{||}, E)g_{N,1}(k_{||}, z_+)], \tag{8}$$

where,

$$B_p(k_{||}, E) = i\,[\Gamma_p(k_{||}, z_+) - \Gamma_p(k_{||}, z_-)], \tag{9}$$

with $p = 1, N$, $z_\pm = E \pm i\varepsilon$ and

$$\Gamma_p(k_{||}, E) = \begin{cases} S_{1,0}(k_{||})G_L^{sf}(k_{||}, z)S_{0,1}(k_{||}); & p = 1 \\ S_{N,N+1}(k_{||})G_R^{sf}(k_{||}, z)S_{N+1,N}(k_{||}); & p = N. \end{cases} \tag{10}$$

The surface Green's functions of the electrodes $G_{L,R}^{sf}(k_{||}, z)$ are constructed scalar-relativistically, which allows us to decompose the conductance into spin-conserving and spin-flip components [19]. $g_{1,N}$ and $g_{N,1}$ are the upper right corner and lower left corner components of the auxiliary Green's function matrix

$$g = [P\text{-}S]^{-1}. \tag{11}$$

Here, $P$ is the fully relativistic potential function (6) and $S$ the tridiagonal matrix of scalar relativistic structure constants [15].

The heterostructure is 'grown' in the [100] direction. We have tried two different geometries: one where the left semi-infinite GaAs is separated from the right semi-infinite GaAs by 4 monolayers of AlAs and another where the separation is 8 monolayers. The self-consistent charge distribution is achieved within scalar-relativistic TB-LMTO calculations for GaAs/AlAs heterostructures treated using supercells with 6 monolayers of GaAs separated by the number of AlAs monolayers that correspond to the above mentioned geometries. The transmission is calculated at zero bias with lateral $k_\parallel$ taken in the [001] direction.

In GaAs we included only Ga 4d orbitals, which provide a slightly larger fundamental band gap than it is normally predicted within LDA (when Ga 3d orbitals are included in the basis). The important part of the calculated band structures of GaAs and AlAs are shown in Fig. 1. The GaAs direct band gap is 0.53 eV compared to the experimental value of 1.52 eV, the AlAs direct ($\Gamma$-$\Gamma$) and indirect (fundamental) gaps are 2.22 eV and 1.25 eV, respectively, compared to the experimental values of 3.13 eV and 2.23 eV, respectively. We would like to point out that the energies examined in this work are in the vicinity of conduction band minimum. Even though the band gaps are smaller than the experimental they are big enough to ensure that the complex band structure around conduction minimum has the correct character [19]. Based on this justification, LDA based transport calculations through heterostructures that include semiconductors have been used before [20] and revealed important information that is impossible to obtain from a model calculation. The effective mass of incoming electrons is smaller than the experimental value and this is an inherent problem of the LDA; to this date the proper treatment of electron-electron correlations in ab-initio transport methods has not been addressed and to our knowledge and experience it is a very difficult problem on it's own. In fact, even the most sophisticated techniques, like GW ($G$ = Green's function, $W$ = screened Coulomb interaction) approximation, cannot match the experimental effective mass and ad-hoc adjustments are required [21]; moreover these sophisticated techniques require heavy computational resources hence they are used only in bulk geometry. The use of such techniques in layered or supercell geometry is currently out of question. Also, *ad-hoc* techniques are known to have difficulty in matching simultaneously both the band gap and effective mass to the experiment. For example, in Ref. [22], while the band gap of GaAs is correct, the conduction band minimum effective mass is off by 33%. In the problem we are currently examinining the GaAs electron effective mass controls only the width of the Fano resonances. As we will explain straightaway, several other parameters that are related to our problem are predicted well. The valence band offset is 0.36 eV in good agreement with previously calculated from the individual bulk materials [23]. The conduction band offset ($\Gamma$-$X$) is 0.37 eV. Many important features of the tunneling relevant to our discussion are controlled by the conduction band offset and the Dresselhaus splitting in the vicinity of X point of AlAs [3]. In the heterostructure there is a charge transfer that increases the valence band offset to 0.53 eV and therefore we obtain a value of 0.2 eV for the conduction band offset, which is very close to the value used in our empirical tight-binding calculations (0.16 eV) [3]. The coefficient $\beta$ for the linear Dresselhaus splitting at $X$ point ($\Delta E = \beta k_\parallel$)comes out at 0.108 eVÅ. There are no available experimental estimates for $\beta$. In Ref. [21] the LDA bands and Dresselhaus splittings where compared to the predictions of the 'scaled' Quasiparticle self-consistent $GW$ (QS$GW$) method which gives very accurate predictions of band eigenvalues and eigenvectors. It was found that both methods agree very well at the $X$ point. Not surprisingly, the agreement is also good for the transverse and longitudinal effective masses for AlAs. The LDA values for transverse and longitudinal masses are 0.232 $m_0$ and 0.836 $m_0$, respectively, while the QS$GW$ method predicts 0.225 $m_0$ and 0.738 $m_0$, respectively. While $\beta$ dictates the magnitude of the spin splitting of the quasi-bound levels at $X$, the effective masses together with the barrier width dictate their energy position (hence, separation). Therefore, despite the apparent shortcoming of the LDA to give an overall reliable prediction of the conduction band structure of the materials involved here, separate aspects that are important in the process of indirect tunneling are predicted well.

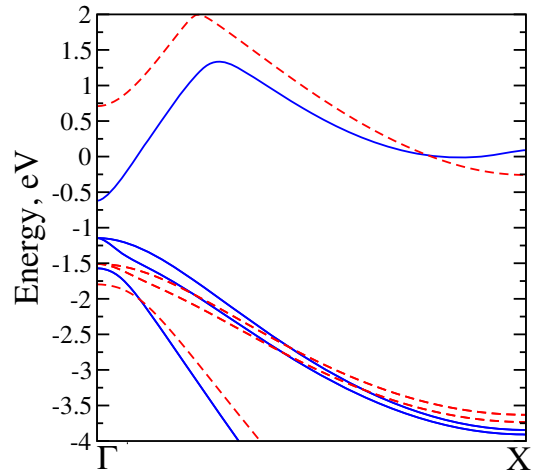

**Figure I**
Fully-relativistic LMTO-ASA within LDA band structures of AlAs and GaAs. The band structures of AlAs and GaAs are plotted with dashed red line and solid blue line, respectively. The valence band offset is 0.36 eV, the GaAs direct band gap is 0.53 eV, the AlAs direct ($\Gamma$-$\Gamma$) and indirect (fundamental) gaps are 2.22 eV and 1.25 eV, respectively. The conduction band offset ($\Gamma$-$X$) is 0.37 eV.

In Fig. 2 we show the transmission for a heterostructure with 4 monolayers of AlAs. For $k_{||}$ = 0.03 $2\pi/a$ the tunneling starts at approximately -0.4 eV (for $k_{||}$ = 0 it would start at approximately -0.5 eV). Below -0.3 eV the tunneling transmission exhibits a typical direct tunneling behavior, and the spin polarization is insignificantly small. Between -0.3 and -0.2 eV the tunneling electron experiences the first Fano resonance-antiresonance due to the interaction with the discrete levels in the $X$ valley well; the resonances and zeros for the two spin channels occur at slightly different energies, resulting in the sharp peaks of spin polarization seen in the second panel of Fig. 2. A second resonance-antiresonance occurs just below 0.05 eV, resulting in another spin polarization 'hot spot'. Generally, for the given values of $k_{||}$ and barrier size we find only 'hot spots' of spin polarization in the vicinity of Fano resonances, but no energy windows of large spin polarization. In Fig. 3 we show the transmission for the same values of $k_{||}$ but for a heterostructure with 8 monolayers of AlAs. For $k_{||} = 0.03 \ 2\pi/a$, there is a 0.05 eV energy window, around -0.2 eV, where the spin polarization of the transmission is approximately 25%, but no such window is observed for $k_{||} = 0.06 \ 2\pi/a$. The Dresselhaus splitting is bigger for the latter value of $k_{||}$, but the spin polarization of the transmission coefficient is smaller. This non-trivial behavior was predicted by the empirical tight-binding model that we used previously: the well separated resonances below and above -0.2 eV for $k_{||} = 0.03 \ 2\pi/a$ make the polarization window relatively wide [3]. As it was predicted by Ref. [3], wider barriers provide better conditions for well separated resonances and a proper order of corresponding antiresonances. In contrast, for $k_{||} = 0.06 \ 2\pi/a$ the transmission coefficient is very different due to nonparabolicity of the bands, (parabolic bands would provide more or less similar transmission coefficient for different $k_{||}$) thereby the spin polarization is less efficiently filtered in the barrier. The same pattern is also observed in Fig. 2, i.e., there are very different transmission coefficients for different values of $k_{||}$. Therefore, our *ab-initio* calculations show that the full $k_{||}$-dependence should be considered in order to have a more realistic picture of the spin polarization.

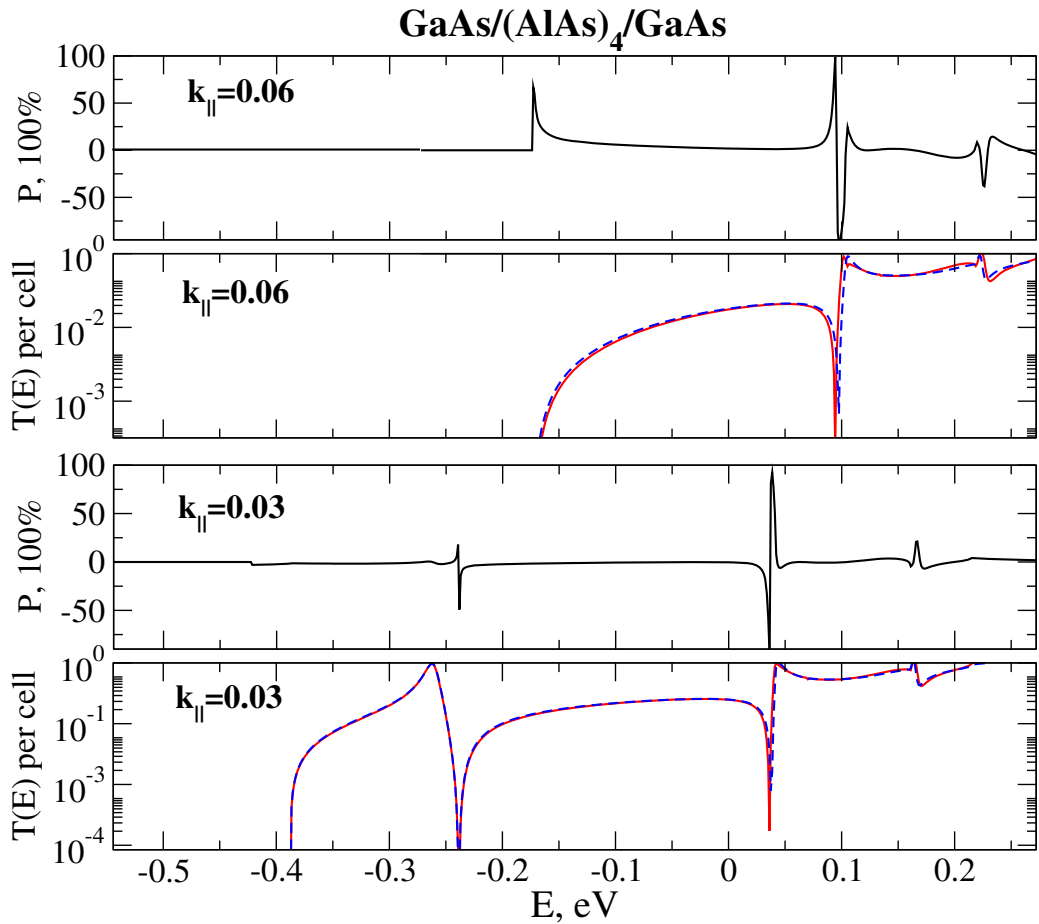

**Figure 2**
The transmission coefficient and its spin polarization for a GaAs/(AlAs)$_4$/GaAs single barrier structure. The red solid line is for 'spin-up' and the blue dashed line is for 'spin-down'. Two cases of $k_{||}$ are shown. Hot spots of large spin polarization exist in the vicinity of Fano resonances.

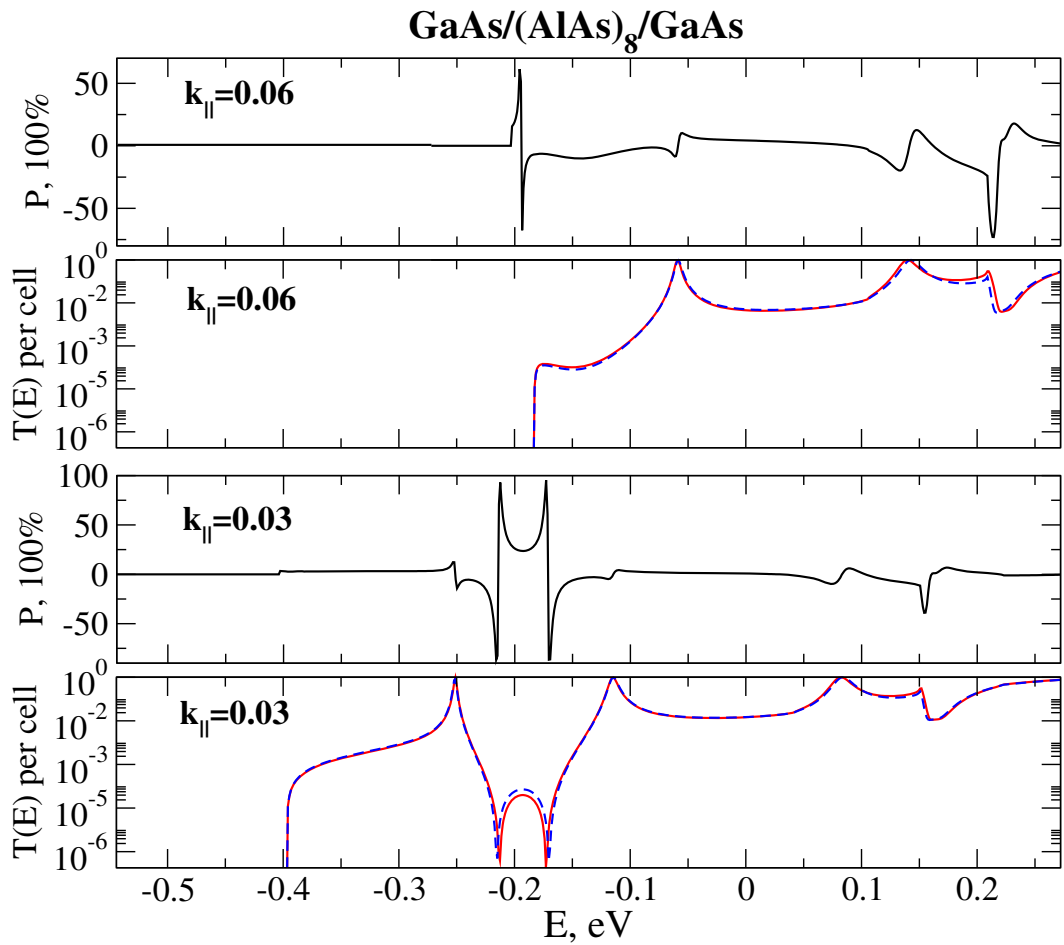

**Figure 3**
The transmission coefficient and its spin polarization for a GaAs/(AlAs)$_8$/GaAs single barrier structure. The red solid line is for 'spin-up' and the blue dashed line is for 'spin-down'. Two cases of $k_{||}$ are shown. One energy window of large spin polarization exists in the vicinity of -0.2 eV for $k_{||}$ = 0.03 $2\pi/a$. No such window is obtained for $k_{||}$ = 0.06 $2\pi/a$. This shows that efficient spin polarization can be obtained independently of the magnitude of Dresselhaus splitting.

We make a comment on the ab-initio method that we use in this paper. The surface Green functions of the left/right semi-infinite GaAs regions are scalar-relativistic, thus the Dresselhaus term is not taken into account in there. We consider also GaAs layers in the active region where the Hamiltonian is fully relativistic, but the transmission is for an electron that starts at the left semi-infinite and ends at the right semi-infinite region. However, we already questioned the importance of the Dresselhaus term in contacts [24]. The related calculations [25] show that this term has a minor effect on the overall spin polarization with a much greater role played by the current operator. In our ab-initio calculations the current operator is properly taken into account.

Finally, we comment about experimental aspects of the present subject. An experimental setup would measure currents with contributions from all other energy and momentum values that, in principles, will wash out the spin polarization. For instance, electronic states with the same energy, corresponding to $k_{||}$ and $-k_{||}$, will have opposite spins, thus the total current will carry the same amount of spin-up and spin-down. To fix these aspects we already suggested [3] that the electrons should be injected from a resonant tunneling diode in order to focus the electrons in the $k_{||}$-plane [26]. Moreover, by applying a voltage bias in the $k_{||}$-plane of the emitter [27,28], the isotropy of $k_{||}$ will be further broken, thus a net spin polarization will be induced in the energy window.

## Conclusion

In conclusion, we applied a fully-relativistic first-principles transport method to the spin-dependent tunneling through a GaAs/AlAs/GaAs [100] heterostructure. The method is based on the Green's function representation of the Tight-Binding Linear Muffin-Tin Orbital basis in the Atomic Spheres Approximation. The calculations were performed in the Local Spin Density Approximation within the Density Functional Theory.

Considering the full band structure of the GaAs/AlAs/GaAs [100] system, these calculations confirm previous general predictions made with a simplified empirical tight-binding method. Namely, in order to have windows with large spin polarizations, two conditions need to be satisfied: the first is to have well separated resonances such that their corresponding anti-resonances do not interact with each other and the second is that the energy order of the resonances in the spin channels have to be the same as the energy order of their corresponding zeros.

We found a large energy window of 25–50% spin polarization for a barrier of 8 AlAs monolayers at $k_{||} = 0.03 \ 2\pi/a$, but there is no such energy window at $k_{||} = 0.06 \ 2\pi/a$. Our study suggests that, in order to find energy windows with large spin polarization, a detailed knowledge of the energy dependence on $k_{||}$ and spin must be considered.

## Acknowledgements

The authors gratefully acknowledge the financial support from the Office of Naval Research and from Romanian Ministry of Education and Research.